# Difference in the wind speeds required for initiation versus continuation of sand transport on Mars: Implications for dunes and dust storms

Jasper F. Kok[1,2,*]

[1]Department of Atmospheric, Oceanic, and Space Sciences, University of Michigan, Ann Arbor, Michigan, USA.
[2]Present address: Advanced Study Program, National Center for Atmospheric Research, Boulder, Colorado, USA

**ABSTRACT**

Much of the surface of Mars is covered by dunes, ripples, and other features formed by the blowing of sand by wind, known as saltation. In addition, saltation loads the atmosphere with dust aerosols, which dominate the Martian climate. We show here that saltation can be maintained on Mars by wind speeds an order of magnitude less than required to initiate it. We further show that the resulting hysteresis effect causes saltation to occur for much lower wind speeds than previously thought. These findings have important implications for the formation of dust storms, sand dunes, and ripples on Mars.

ARTICLE

*Introduction.-* Wind-blown sand, also known as saltation, has shaped the Martian surface by creating dunes, ripples, and a plethora of erosional features [1-4]. Moreover, saltation occurring in dust storms and dust devils [5-7], which are common on Mars [8], loads the atmosphere with mineral dust, the radiative effects of which shape Martian weather and climate [6,7].

Saltation occurs on present-day Mars, as shown by the accumulation of sand on the deck of the Spirit rover [3], the seasonal reversal of dune slip faces in Proctor Crater [9], and the movement of sand dunes [9,10] and ripples [4] on the Martian surface. It is therefore puzzling that both lander measurements [6,11] and atmospheric circulation models [9,12] indicate that wind speeds on Mars rarely exceed the threshold required to initiate saltation. For example, atmospheric simulations in a region of seasonally shifting dune slip faces found hourly-averaged wind speeds to reach a maximum of only about a third of the saltation threshold [9].

The characteristics of Martian saltation are thus poorly understood, partially due to the inherent difficulty of obtaining experimental data either in wind-tunnels [1,5,13] or on the Martian surface [2-4]. Numerical models are therefore an especially valuable tool for studying the characteristics of Martian saltation [13-15].

Early simulations of saltating particle trajectories were performed by White and co-workers [13], who reported Martian trajectories somewhat larger than on Earth. However, the more detailed numerical study of Almeida et al. [14] found that saltation trajectories on Mars are approximately two orders of magnitude higher than on Earth. This study simulated essential features of saltation, but did not explicitly include the splashing [16] of surface particles by impacting saltating particles. However, splashing determines many of the characteristics of saltation [16-18]. For example, it determines the "impact threshold" [17,19], which denotes the lowest shear velocity (defined as $u^* = \sqrt{\tau/\rho_a}$, where $\tau$ is the wind shear stress and $\rho_a$ is the air density) at which saltation, after being initiated, can be sustained by the splashing of surface particles.

*Numerical saltation model (COMSALT).-* Here we improve upon the above studies by investigating Martian saltation with a comprehensive physically-based numerical model of steady-state saltation, called COMSALT, that has been extensively tested with terrestrial measurements [17]. COMSALT explicitly simulates the trajectories of saltating particles due to gravitational and fluid forces, and accounts for the stochasticity of individual particle trajectories due to turbulence and collisions with the irregular soil surface. Moreover, the retardation of the wind by the drag of saltating particles is explicitly simulated. Finally, in contrast to most previous models [13-15], COMSALT includes a physically-based parameterization of the splashing of surface particles, based on conservation of energy and momentum. This parameterization is in good agreement with a variety of laboratory experiments, including measurements of the number and speed of ejected particles [17]. Because of the explicit inclusion of splash, as well as other improvements over previous studies [13-16], COMSALT is the first numerical model capable of reproducing terrestrial measurements of the impact threshold, as well as a wide range of other measurements [17]. Here, we for the first time apply COMSALT to Mars, using an air pressure and temperature of 700 Pa and 220 K, a gravitational acceleration of 3.72 m/s$^2$, and a particle density of 3000 kg/m$^3$ [7,20].

*Results.-* Surprisingly, we find that the impact threshold required to maintain saltation on Mars is an order of magnitude smaller than the "fluid threshold" required to initiate it (Fig. 1). The ratio of impact to fluid thresholds on Mars is thus much smaller than the value of ~0.82 observed for loose sand on Earth [19]. The difference occurs because the lower gravity and vertical drag on Mars makes particles travel higher and longer trajectories [14], causing them to be accelerated by wind for a much longer duration during a single hop than on Earth. This effect combines with the lower atmospheric density on Mars to produce an impact threshold that is comparable to that on Earth (Fig. 1). Since the lower atmospheric density on Mars causes the fluid threshold there to be an order of magnitude larger than on Earth [1], the resulting ratio of the impact to fluid thresholds is thus much smaller on Mars than it is on Earth (Fig. 1).

In addition to numerical simulations, we recently performed an analytical analysis of the impact threshold that yielded a similar result (see Ref. [21] and Fig. 1). Moreover, two previous studies [14,20] also suggested that the ratio of impact to fluid thresholds on Mars is smaller than on Earth. Finally, indirect experimental support for the low Martian impact threshold is the puzzling occurrence of ripples and dune-like bedforms of ~100 µm particles on Mars [2,3,20,22]. Such light particles are thought to be easily suspended by turbulence at the fluid threshold [1,23] (Fig. 1), such that the wind speed at which these bedforms develop must be substantially below the fluid threshold. The occurrence of these bedforms on the Martian surface thus requires the impact threshold to be substantially smaller than the fluid threshold, as we predict (Fig. 1).

The small ratio of the impact and fluid thresholds allows Martian saltation to occur for much lower wind speeds than previously thought possible [24]. Indeed, once saltation is initiated by a localized wind gust, it will continue downwind until the wind speed falls by approximately an order of magnitude to a value below the impact threshold (Fig. 1). The occurrence of saltation transport at instantaneous wind speeds between the impact and fluid thresholds thus depends on the history of the system, a phenomenon known as "hysteresis." Specifically, saltation transport will be initiated when the instantaneous wind speed exceeds the fluid threshold, and will be halted when the instantaneous wind speed drops below the impact threshold.

We quantify the effect of hysteresis by determining the probability that saltation transport takes place at any given moment as a function of the mean wind shear velocity and the impact and fluid thresholds. To do so, we assume that the probability distribution of the instantaneous wind speed can be described by the Weibull distribution [25]. The probability that the instantaneous wind speed $U$ at a given reference height exceeds a value $U_r$ is then approximately [25]

$$P_W(U > U_r) = \exp\left\{-\left[\frac{U_r}{\overline{U}}\Gamma\left(\frac{k+1}{k}\right)\right]^k\right\}, \tag{1}$$

where $\overline{U}$ is the average wind speed, $\Gamma$ is the gamma function, and the shape factor $k$ has a value between 1.5 and 3; we assumed $k = 2$ [25]. The probability $P_{tr}$ that transport occurs is then the sum of (i) the probability that the instantaneous wind speed exceeds the fluid threshold, and (ii) the probability that the wind speed is between the fluid and impact thresholds, and that the wind speed exceeded the fluid threshold more recently than that it dropped below the impact threshold. That is:

$$P_{tr} = P_W(U > U_{ft}) + \frac{P_W(U_{it} < U < U_{ft})P_W(U > U_{ft})}{P_W(U < U_{it}) + P_W(U > U_{ft})}, \qquad (2)$$

where $U_{it}$ and $U_{ft}$ are the instantaneous wind speeds corresponding to the impact ($u*_{it}$) and fluid ($u*_{ft}$) thresholds. By combining (1) and (2), we obtain

$$P_{tr} = \exp\left[-\left(u*_{ft}\Gamma(1+1/k)/\overline{u*}\right)^k\right]\left\{1 + \frac{\exp\left[-\left(u*_{it}\Gamma(1+1/k)/\overline{u*}\right)^k\right] - \exp\left[-\left(u*_{ft}\Gamma(1+1/k)/\overline{u*}\right)^k\right]}{1 - \exp\left[-\left(u*_{it}\Gamma(1+1/k)/\overline{u*}\right)^k\right] + \exp\left[-\left(u*_{ft}\Gamma(1+1/k)/\overline{u*}\right)^k\right]}\right\}, \qquad (3)$$

where $\overline{u*}$ is the mean shear velocity averaged over a sufficiently long time period [25], and where we used that $\overline{U} \propto \overline{u*}$ [e.g.,1,17].

The hysteresis of Martian saltation thus allows sand transport to occur at shear velocities well below the fluid threshold, with the probability that transport occurs given by Eq. (3). To illustrate this phenomenon, we use Eq. (3) to calculate sand transport at Proctor Crater, where seasonally-shifting dune slip faces indicate regularly occurring saltation. This is puzzling, since a mesoscale atmospheric simulation of Proctor Crater found the maximum hourly-averaged wind shear velocity to be only about a third of the fluid threshold [9] (Fig. 2). Nonetheless, there is a small but finite probability that the instantaneous wind speed will exceed the fluid threshold and thus initiate saltation. Following initiation, saltation will be maintained for a relatively long period of time because of the low value of the impact threshold (Fig. 1). As a result, Eq. (3) predicts a substantial probability of saltation transport at Proctor Crater for hourly-averaged wind shear velocities that are on the order of the impact threshold, and thus well below the fluid threshold (Fig. 2). This result helps explain the observation that dune slip faces at Proctor Crater shift seasonally, even though the hourly-averaged shear velocity remains well below the fluid threshold [9].

*Discussion.-* Our finding that saltation occurs for shear velocities well below the fluid threshold has potentially important implications for the emission of dust aerosols on Mars. While some recent studies indicate that dust emission through the lifting of aggregates of air-fall dust likely occurs before the saltation fluid threshold is reached [4,26], the hysteresis of Martian saltation allows dust emission through saltation bombardment [5] to also occur well below the fluid threshold (see Eq. (3) and Fig. 2). Additional research is thus required to determine the relative importance of both processes in supplying the Martian atmosphere with mineral dust.

In addition to allowing sand transport and dust emission to occur well below the fluid threshold, hysteresis also affects the characteristics of Martian saltation. In particular, the typical wind speed experienced by saltating particles is substantially lower than suggested by the fluid threshold [27]. As a consequence, particle speeds and the size of particle trajectories are substantially smaller than found by a previous model [14] that did not include splash (see Fig. 3). Our results are supported by the simple theoretical argument that saltating particles must have an average impact speed that causes the ejection or rebound of one particle per collision in steady state [16-18,21]. This average impact speed likely depends on both the gravitational acceleration and the characteristics of the soil bed and equals ~1 – 1.5 m/s for loose sand on Earth [17]. On Mars, the reduced gravity and increased particle size [9,23] combine to produce an average impact speed that is comparable to that observed for Earth conditions (see Fig. 3a and Ref. [21]). Note that these arguments predict that the average particle speed at the surface, which depends on both impacting and rebounding particles, stays approximately constant with shear velocity [16,17]. This has indeed been confirmed by recent wind-tunnel experiments [28,29] (Fig. 3a), which provides additional experimental support for our results.

The smaller size of saltation trajectories has implications for the minimal size of Martian dunes, which depends on the horizontal length that is required for the sand flux to reach its saturated value [30-33]. This saturation length depends on the length of saltation trajectories [30,31], which in turn scales with the drag length required to accelerate particles to the wind speed [32,33]. But a simple scaling using the drag length predicts a minimal size of Martian crescent-shaped barchan dunes an order of magnitude larger than observed [31,33]. Parteli et al. [31] hypothesized that this discrepancy is due to much larger splashing events in Martian saltation, which would decrease the saturation length and therefore the dune size. However, we find that splashing events in saltation on Mars are similar to those on Earth because the impact speed is comparable (see Fig. 3a and Ref. [21]). Instead, our results indicate that the previous overestimation of the minimal size of Martian dunes [31,33] can be resolved by accounting for the low value of the Martian impact threshold, which results in much smaller saltation trajectories [see Fig. 3(b) and Ref. [27]], and thus smaller dunes, than found by previous studies [14,31]. Indeed, we find saltation trajectories that are only an order of magnitude larger than on Earth (Fig. 3b), in approximate agreement with the scaling of the minimal size of Martian (~100 meters) to terrestrial (~10 meters) barchan dunes [31]. A detailed dune model [30,31] is needed to test these preliminary estimates.

*Conclusion.-* We used numerical simulations to show that a substantial hysteresis effect occurs in Martian saltation (Fig.1), which causes saltation to take place for much lower wind speeds than previously thought [1]. This finding helps resolve the various observations of active saltation [3,4,9,10] with measurements [6,11] and atmospheric simulations [9,12] that show that surface wind speeds rarely exceed the saltation threshold (Fig. 2). Moreover, the hysteresis effect could allow dust emission through saltation bombardment to occur for much lower wind speeds, with important implications for the formation of dust storms and dust devils and thus the Martian climate [6,7]. Finally, accounting for the effect of hysteresis results in much smaller saltation trajectories (Fig. 3) than previously calculated [14], which potentially resolves the mismatch between predictions and observations of the minimal size of Martian barchan dunes [31,33]. The hysteresis of Martian saltation thus has implications for a wide range of geological and atmospheric processes on Mars.

This work was supported by the National Science Foundation (grant ATM 0622539) and the National Aeronautics and Space Administration (grant NNX07AM99G). Comments by Shanna Shaked, Nilton Renno, and two anonymous reviewers improved the paper.

**REFERENCES**

*Electronic address: jfkok@umich.edu

[1] R. Greeley and J. D. Iversen, *Wind as a Geological Process* (Cambridge Univ. Press, Oxford, 1985).
[2] R. Sullivan *et al.*, Nature **436**, 58 (2005).
[3] R. Greeley *et al.*, J. Geophys. Res. **111**, E02S09 (2006).
[4] R. Sullivan *et al.*, J. Geophys. Res. **113**, E06S07 (2008).
[5] R. Greeley, Planet. Space Sci. **50,** 151 (2002).
[6] R. Zurek *et al.* in *Mars*, edited by H. H. Kieffer *et al.* (Univ. Arizona Press, Tucson, 1992), pp. 835-933.
[7] C. Leovy, Nature **412**, 245 (2001).
[8] B. A. Cantor *et al.*, J. Geophys. Res. **106**, 23653 (2001).
[9] L. K. Fenton, A. D. Toigo, and M. I. Richardson, J. Geophys. Res. **110**, E06005 (2005).
[10] M. C. Bourke, K. S. Edgett, and B. A. Cantor, Geomorphology **94**, 247 (2008).
[11] R. Sullivan *et al.*, J. Geophys. Res. **105**, 24547 (2000).
[12] R. M. Haberle, J. R. Murphy, and J. Schaeffer, Icarus **161**, 66 (2003).
[13] B. R. White *et al*., J. Geophys. Res. **81**, 5643 (1976); B. R. White, J. Geophys. Res. **84**, 4643 (1979).
[14] M. P. Almeida *et al*., Proc. Natl. Acad. Sci. USA **105**, 6222 (2008).
[15] J. F. Kok and N. O. Renno, Geophys. Res. Lett. **36**, L05202 (2009); Phys. Rev. Lett. **100**, 014501 (2008).
[16] J. E. Ungar and P. K. Haff, *Sedimentology* **34**, 289 (1987).
[17] J. F. Kok and N. O. Renno, J. Geophys. Res. **114**, D17204 (2009).
[18] B. Andreotti, *J. Fluid Mech.* **510**, 47 (2004).
[19] R. A. Bagnold, *Geograph. J.* **89,** 409 (1937); J. D. Iversen, and K. R. Rasmussen, Sedimentology **41**, 721 (1994).
[20] P. Claudin and B. Andreotti, *Earth Planet. Sci. Lett.* **252**, 30 (2006).
[21] J. F. Kok, arXiv: 1001.4840
[22] R. L. Fergason *et al.*, J. Geophys. Res. **111**, E02S21 (2006).
[23] K. S. Edgett and P. R. Christensen, J. Geophys. Res. **96**, 22765 (1991).
[24] Note that COMSALT [17] and Ref. [21] assume loose sand, typical for active dunes and ripples. Our results thus do not apply to cemented sandy deposits, which are usually inactive [2-4]. Also note that the assumption of loose sand causes an underestimation of the impact threshold of particles smaller than ~100 – 150 μm [19,21], for which cohesive forces can exceed the gravitational force [1,26].
[25] C. C. Wu, and A. N. L. Chiu, J. Wind Eng. Ind. Aerodyn. **13**, 43 (1983); J. V. Seguro and T. W. Lambert, J. Wind Eng. Ind. Aerodyn. **85**, 75, (2000).
[26] J. P. Merrison *et al.*, Icarus **191**, 568 (2007).
[27] See supplementary material at http://link.aps.org/supplemental/10.1103/PhysRevLett.000.000000 for the simulated wind profile in Martian saltation.
[28] K. R. Rasmussen, and M. Sorensen, J. Geophys. Res. **113**, F02S12 (2008).
[29] M. Creyssels *et al.*, J. Fluid Mech. **625**, 47 (2009).
[30] G. Sauermann, K. Kroy, and H. J. Herrmann, Phys. Rev. E **64**, 031305 (2001).
[31] E. J. R. Parteli and H. J. Herrmann, Phys. Rev. E **76**, 041307 (2007).

[32] P. Hersen, S. Douady, and B. Andreotti, Phys. Rev. Lett. **89**, 264301 (2002).
[33] K. Kroy, S. Fischer, and B. Obermayer, J. Phys.: Condens. Matter **17**, S1229 (2005).
[34] S. L. Namikas, Sedimentology **50**, 303 (2003).

**FIGURES**

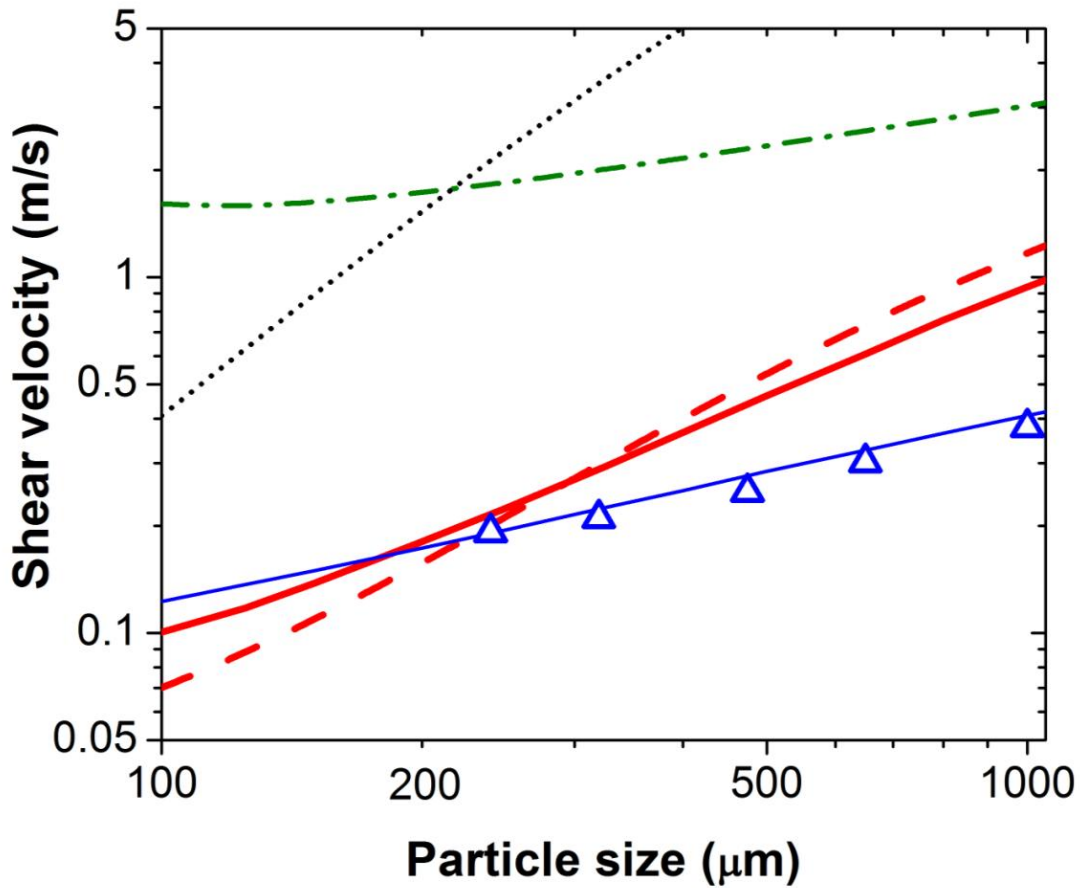

FIG. 1 (Color online). Saltation impact threshold on Earth and Mars. The impact threshold simulated with COMSALT [17] for Earth (thin blue solid line) matches measurements [19] (triangles). On Mars, both numerical simulations with COMSALT [17] (thick red solid line) and recent analytical calculations [21] (thick red dashed line) show that the impact threshold is approximately an order of magnitude below the fluid threshold [1] (dash-dotted green line). Also included is the approximate shear velocity at which particles become suspended (dotted black line; obtained from $u^* \approx v_t$ [1,23], where $v_t$ is the particle's terminal velocity).

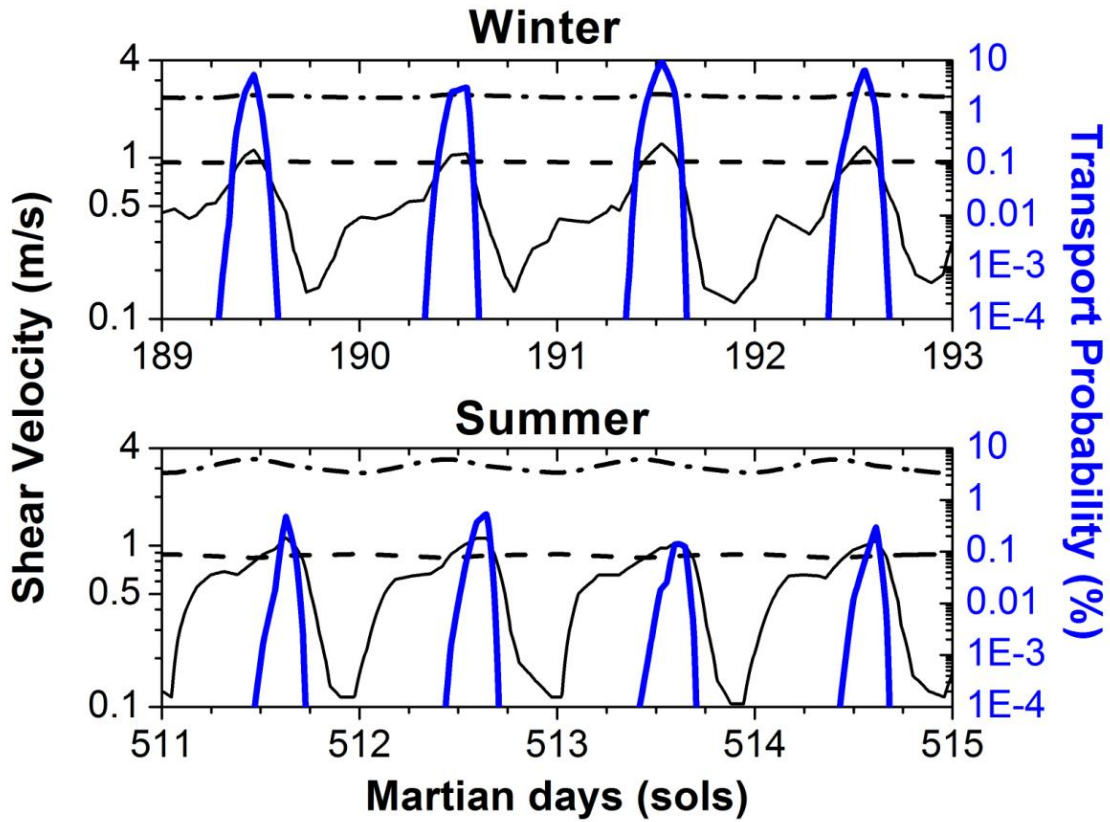

FIG. 2 (Color online). Seasonally-reversing dune slip faces indicate active saltation at Proctor Crater, but hourly-averaged wind shear velocities simulated by a mesoscale model [9] (thin black solid line) are substantially below the fluid threshold [1] (dash-dotted line) for both southern hemisphere winter (top) and summer (bottom). Sand transport nonetheless occurs because the hysteresis of Martian saltation (see text) causes the probability that saltation transport takes place at any given moment (thick blue solid line and right axis; see Eq. (3)) to be substantial when the hourly-averaged wind shear velocity is on the order of the impact threshold (dashed line; from Ref. [21]; assumed particle size was 740 μm [9]).

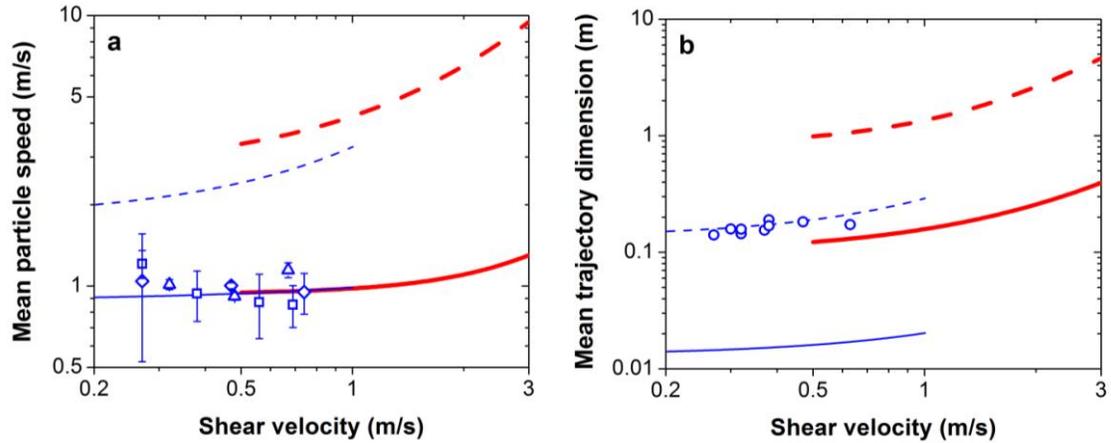

FIG. 3 (Color online). Mean particle speed and hop dimensions for saltation on Earth and Mars. Thin blue lines and symbols represent terrestrial predictions and measurements, while thick red lines denote predictions under Martian conditions. (a) Average horizontal particle speed at the surface (solid lines) and particle speed averaged over all saltating particles (dashed lines) calculated by COMSALT for Earth (250 μm particles, typical of terrestrial dunes [1]) and Mars (500 μm particles, typical of Martian dunes [9,23]). The symbols denote measurements of the horizontal particle speed linearly extrapolated to the surface (see Fig. 5 of Ref. [29]; we used measurements within 2 mm of the surface) of respectively 242 μm (squares [28] and triangles [29]) and 320 μm particles (diamonds [28]). Error bars were derived from the uncertainty in the fitting parameters. (b) The mean height (solid lines) and length (dashed lines) of saltation trajectories as a function of shear velocity for Earth and Mars. Included for comparison is the average deposition length [34] in field measurements (circles), which is a good measure of the average saltation length and was obtained from Fig. 8a of Ref. [34].

10